\documentclass[twocolumn,prb,showpacs,multicol,amsmath,amssymb]{revtex4}
\usepackage[dvips]{graphicx}

\begin{document}
\def\be{\begin{equation}}
\def\ee{\end{equation}}
\def\bea{\begin{eqnarray}}
\def\eea{\end{eqnarray}}
\def\no{\nonumber}
\def\ba{\begin{array}}
\def\ea{\end{array}}
\title{The effect of Aharanov-Bohm phase on the 
magnetic-field dependence of two-pulse echos in glasses at low
    temperatures}
\author{A. Akbari$^1$ and A. Langari$^{1,2}$}
\affiliation{
$^1$Institute for Advanced Studies in Basic Sciences,
P. O. Box 45195-1159, Zanjan, Iran\\
$^2$Max-Planck-Institut f\"ur Physik komplexer Systeme, N\"othnitzer Str.38,
01187 Dresden, Germany
}
\date{\today}
\begin{abstract}
The anomalous response of glasses in the echo
amplitude experiment is explained in the presence of a magnetic field. We have
considered the low energy excitations in terms of an effective two
level system. The effective model is constructed on the flip-flop
configuration of  two interacting two level systems. The magnetic
field affects the tunneling amplitude through the Aharanov-Bohm
effect. The effective model has a lower scale of energy in
addition to the new distribution of tunneling parameters which
depend on the interaction. We are able to explain some features 
of echo amplitude versus a magnetic field, namely, the dephasing
effect at low magnetic fields, dependence on the strength of
the electric field, pulse separation effect and the influence of temperature.
However this model fails to explain the isotope effects which essentially
can be explained by the nuclear quadrupole moment.
We will finally discuss  the features of our
results.

\end{abstract}
\pacs{61.43.Fs  64.90.+b  77.22.Ch  72.20.Ht  }

\maketitle

%%%%%%%%%%%%%%%%%%%%%%%%%%%%%%%%%%%%%%%%%%%%%%%%%%%%%%%%%%%%%%%%%%%%%%%%%%%%%%%

\section{introduction}

Unusual behavior of the heat capacity and thermal conductivity
of glasses at very low temperatures \cite{zeller} can be governed
by a broad distribution of low-energy excitations.
These could be described as two level tunneling systems whose
energies  and tunneling amplitudes are randomly distributed
\cite{phill_72, ander_72}. The origin of these excitations
is assumed to come from atoms or group of atoms that can
be arranged in two energetically close configurations.
Such a configuration can be a cluster of few hundred atoms or
molecules \cite{lubchenko2001}.
Although the nature of the tunneling entities is obscure, many
theories are based on such a tunneling model to describe the
low temperature properties of glasses.
A two level system (TLS) is characterized by two parameters
of a double well potential, the asymmetry energy $\Delta$,
and the tunneling amplitude $\Delta_0$. The Hamiltonian of
a single TLS in the pseudo-spin operators is given by
\begin{equation}
H_0=\frac{1}{2}(\Delta\sigma_{z}+\Delta_{0}\sigma_{x}).
\label{h0}
\end{equation}
As a consequence of the irregular
structure of glasses, these  two parameters  are widely
distributed and are independent of each other with a uniform
distribution of
\begin{equation}
f^{(1)}(\Delta,\Delta_{0})=\frac{f_0}{\Delta_{0}},
\label{f0}
\end{equation}
where $f_0$ is a normalization factor.
While many experiments can be explained satisfactorily
by this model it fails to explain some results at very
low temperatures, particularly in the presence of a magnetic
field \cite{enss2002}.

It was a general belief that the dielectric properties of
nonmagnetic insulating glasses are independent of the magnetic field.
Some experiments during the last few years opened the question of
magnetic field dependence. The dielectric constant of several
multicomponent glasses like $BaO-Al_2O_3-SiO_2$ shows a surprising
dependence on a weak magnetic field ($\sim 20 \mu T$) at very low
temperatures ($\sim 2 mK$) \cite{strehlow98,strehlow00}. Increasing
the magnetic field the dependence shows an oscillatory behavior
with a period of a few Teslas.
Moreover the amount of magnetic
impurities present in the {\it nonmagnetic glasses} seems to be
irrelevant and the observed magnetic field effect is strongly
influenced by the applied electric field \cite{wohlfahrt2001}. On
the other hand, recent polarization echo experiments show a strong
and nonmonotonic dependence of echo amplitude on the applied
magnetic field \cite{ludwig02-a,ludwig02-b}. These experiments
give us an insight into the coupling of the tunneling centers to the
magnetic field. Namely, the echo amplitude is
almost independent of the frequency of the electric field which means
that tunneling systems with very different energy splittings
behave the same way \cite{ludwig03}.
Moreover the effect is strongly dependent on
the delay time between two pulses and also on the magnitude of
applied electric field. A surprising outcome of these experiments
is a novel isotope effect observed in different glasses
\cite{nagel04}. The recent results show the important influence of
nuclear quadrupole moments on the observed magnetic field
dependence.

The first theoretical attempt to explain the magnetic field
dependence was based on the Aharanov-Bohm phase
\cite{kettemann99}. In this approach the dielectric permittivity
is given by a generalized TLS where the magnetic field enters in
the quantum mechanical phase of a tunneling charged particle in a
Mexican hat potential. It was assumed that such a generalized  TLS
is the result of collective effects between many tunneling centers
in a mesoscopic size which carry a big amount of charge. The
dipole-dipole interaction between tunneling centers is responsible
for the mentioned collective (coherent) motion  in a weakly
interacting regime \cite{langari02}. 
An evidence for the Aharanov-Bohm effect can be found in
the metallic glasses \cite{zimmerman}.
Another approach which is
also based on the Aharanov-Bohm phase explains the phenomena of
dielectric response in terms of coupled pairs of tunneling centers
\cite{wurger02}. The coupling of two TLSs makes the tunneling path
curved and possessing a flux dependent phase. The most
recent theory is based on the nuclear quadrupole moment of
tunneling entities \cite{wfe02, parshin}. In this model the interaction of
nuclear quadrupole moment with the gradient of the electric field
defines the new scale of energy at small magnetic fields. The
different orientations of the nuclear quadrupole moment in each well
lead to different level splittings which affect the dephasing
mechanism and finally the echo amplitude. This model explains
the echo experiments well, however, it fails to get the right
order of magnitude for the magnetic field effect of dielectric
response \cite{bodea04}. It is claimed that the observed effect in
dielectric response is not a static phenomenon but of dynamical
origin \cite{bodea-private}.

The shortcoming of the isolated TLS model to explain physical
phenomena at low temperatures is usually related to the interaction
between tunneling centers \cite{enss97,hunklinger2000}.
Especially, interaction plays an essential role in determining the
relaxation rates at low temperatures \cite{burin04} which
influences the echo experiment response. In this most
recent work it has been shown how the dipole-dipole interaction can
lead to a relaxation rate faster than phononic counterpart at very
low temperatures. The main idea is based on the delocalization process
of the effective TLSs which are the result of two interacting tunneling
centers \cite{burin98,burin-works}. This is in agreement with the
experimental results on the magnetic field dependence of the dielectric
constant and the effective model proposed in
Ref.[\onlinecite{cochec02}]. Here, the magnetic field dependence
of the dielectric constant is related to the field
dependence of the tunneling
amplitude cutoff which is based on the theory developed earlier \cite{medina92}.
In this theory the localization length increases by adding the
magnetic field which favors the delocalization phenomenon.

Accepting the fact that interaction plays an essential role in the
low temperature physics of glasses we are going to explain the
recent polarization echo experiments in the presence of a magnetic
field. Our approach is based on the main idea presented in
Refs.[\onlinecite{kettemann99,burin04,cochec02}]. The interaction
between two tunneling centers is defined in terms of
flip-flop states which are
considered as a new (effective) TLS.
The magnetic field adds the Aharanov-Bohm phase to the tunneling process.
The effective TLS responds to the electric field of two pulses
by means of renormalized parameters and distribution.
We are able to get most of the features of echo experiments, namely,
strong dependence of the echo amplitude on low magnetic fields,
pulse separation, temperature and electric field effects.
Our results show how one can explain the echo experiments by using
the idea of Aharanov-Bohm phase.

In the next section we define our effective model and its renormalized
parameters. The response of an effective TLS to a two pulse echo is
studied in Sec.\ref{two-pulse}. Then,
in Sec.\ref{results} we will explain the different aspects of our
results to show their correspondence with experimental data.
Finally we summerize our results and discuss  the features of
our approach.

%%%%%%%%%%%%%%%%%%%%%%%%%%%%%%%%%%%%%%%%%%%%%%%%%%%%%%%

\section{Effective TLS}
\label{etls}

Suppose that there are two interacting tunneling
systems. The interaction  between the TLSs can be considered by the
Ising Hamiltonian
\begin{equation}
V=\frac{1}{2}\sum_{i,j}U(R_{ij})\sigma_i^z\sigma_j^z \;;
\hspace{1cm} U(R_{ij})=\frac{U_0}{R_{ij}} \label{interaction}
\end{equation}
where $R_{ij}$ is the distance between two TLSs and $U_0$ is the
characteristic coupling constant.
%%%%%%%%%%%%%%%%%%%%%%%%%%%%%%%%%%%%%%%%%%%%%%%  Figure 1  %%%%%%%%%%%%
\begin{figure}
\includegraphics[width=\columnwidth]{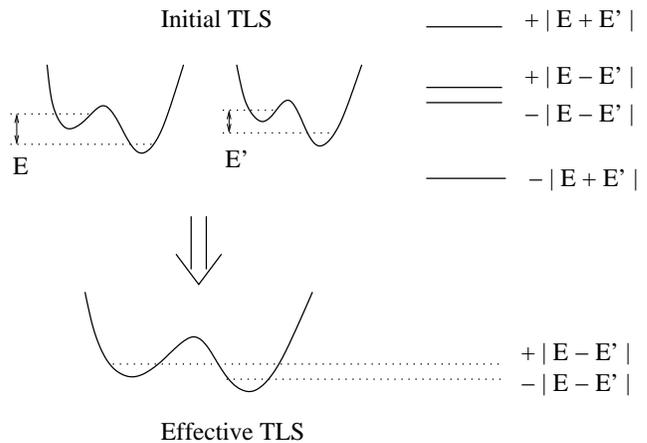}
\caption{Flip-flop configuration as an effective TLS.}
\label{fig1}
\end{figure}
%%%%%%%%%%%%%%%%%%%%%%%%%%%%%%%%%%%%%%%%%%%%%%%%%%%%%%%%%%%%%%%%%%%%%%%

Burin et al.\cite{burin98} proposed that  TLS-TLS coupling in
glasses is dominated at low  temperatures  by a flip-flop
configuration. While the characteristic energy of a single TLS ($E$ or
$E'$, see Fig.(\ref{fig1})) has the same order
of magnitude as temperature ($T$),
the resulting spectrum of two TLSs has two energy scales. The
flip-flop states define a scale of energy which is much less than
$T$ and the other remaining two represent a scale as before, i.e.
$T$. The flip-flop (pair) states correspond to the situation where
one TLS is in the ground state and the other in the excited state.
On the other hand the remaining part of Hilbert space shows both
excited or both grounded TLSs. Concerning the relaxation of the
systems which
leads finally to the dephasing mechanism we only consider the
flip-flop states. Taking into account the interaction between
tunneling centers the Hamiltonian in the flip-flop  subspace is

\begin{equation}
H_p=\frac{1}{2}(\Delta_p\sigma_{z}+\Delta_{0p}\sigma_{x})\nonumber
\end{equation}
where
\bea
\Delta_{0p}=U\frac{\Delta_{01}\Delta_{02}}{E E'}~;~~~~
\Delta_p=\mid E-E'\mid .
\eea

The distribution function for the parameters $\Delta_{0p}$ and
$\Delta_p$ is defined as
\bea\no
f^{(2)}(\Delta_p,\Delta_{0p})=&&<\delta(\Delta_p-\mid
E-E'\mid)\times\\ \label{eq.6}
&&\delta(\Delta_{0p}-\frac{U_0}{R^3}
\frac{\Delta_{01}\Delta_{02}}{E E'})>.
\eea
Therefore by
averaging over the distribution parameters given by Eq.~(\ref{f0})
of the original TLS and ensemble averaging $(\Delta_p\ll E\simeq
T)$ we get
\begin{equation}
f^{(2)}(\Delta_p,\Delta_{0p})\propto\frac{T}{\Delta_{0p}^2}.
%\nonumber
\label{f2}
\end{equation}

Now instead of a  pair of original TLSs we have an effective TLS (ETLS)
which behaves like a single TLS with the new distribution function,
Eq.(\ref{f2}). Such an effective  TLS is aimed to explain the
response of a glass to an electric field of the echo experiment
at low temperatures when a magnetic field is present.
The magnetic field dependence enters via the dependence of the
original tunneling amplitude of pairs
($\Delta_{0i}(\phi)=\Delta_{0i}\cos(\pi \frac{\phi}{\phi_{0}}); i=1,2$) as proposed
in Ref.[\onlinecite{kettemann99}]. By a simple calculation one can
show that the magnetic field dependence of $\Delta_{0p}$ is
similar to the original TLS one's, i.e.
$\Delta_{0p}(\phi)\approx \Delta_{0p}\cos(\pi \frac{\phi}{\phi_{0}})$,
where $\phi_{0}$ is the renormalized quantum of flux ($h/q$). The Planck
constant is denoted by $h$ and the effective charge is $q$.
We should mention that this model is valid while the temperature is
not zero. At the zero temperature the distribution function of ETLSs,
Eq.(\ref{f2}), is zero which means no accessible flip-flop states.
In other words, at $T=0$ the system does not respond to the electric
field of the echo experiments and will remain in the ground state.

%%%%%%%%%%%%%%%%%%%%%%%%%%%%%%%%%%%%%%%%%%%%%%%%%%%%%%%%%%%%%%%%%%%

\section{The response of an effective TLS to a two pulse echo}
\label{two-pulse}

The wave function of an ETLS in an external time dependent electric field is a
linear combination  of the wave function $|\varphi_1\rangle$ and
$|\varphi_2\rangle$ for each level.
\be
\mid \psi\rangle=
a_1(t)e^{\frac{-iE_1t}{\hbar}}\mid\varphi_1\rangle+
a_2(t)e^{\frac{-iE_2t}{\hbar}}\mid\varphi_2\rangle,
\ee
\be\no
\mid
a_1(t)\mid^2 +\mid a_2(t)\mid^2=1.
\ee
%%%%%%%%%%%%%%%%%%%%%%%%%%%%%%%%%%%%%%%%%%%%%%%  figure 2
\begin{figure}[ht!]
\includegraphics[width=\columnwidth]{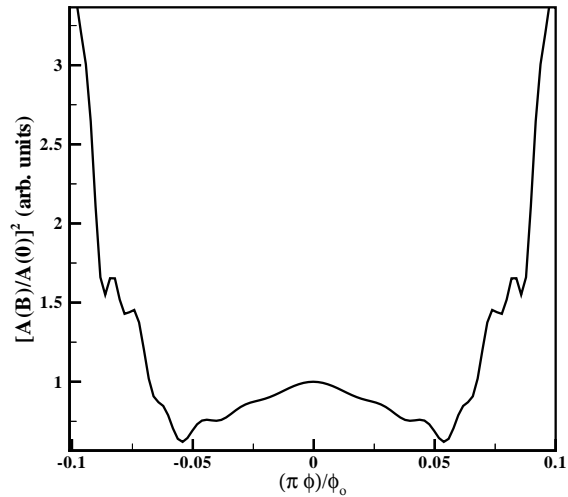}
\caption{The echo amplitude versus the magnetic flux ratio for
relaxation time  $t_0=2\mu s$, by averaging over the effective
distribution function for $0<\frac{\Delta_p}{\hbar}<10^8 Hz $ and
$10^9<\frac{\Delta_{0p}}{2\hbar}<10^9+10^6 Hz $. ($\omega=2\times
10^9 Hz$, $\frac{p_o\xi_o}{2\hbar}=7\times 10^6 Hz $ and
$T=15mK$).} \label{fig2}
\end{figure}
%%%%%%%%%%%%%%%%%%%%%%%%%%%%%%%%%%%%%%%%%%%%%%%%%%%%%%%%%%%%%%%%%%%%%%%

At any time ($t$) during the first pulse of the electric field
($\xi(t)=\xi_o \sin{\omega t}$), the time
variation of the probability amplitudes $a_1(t)$ and $a_2(t)$
is governed by the Schrodinger equation.
In the resonant condition ($\mid\hbar \omega- E\mid\ll \hbar\omega$)
it leads to the following equations\cite{pipard},
imposing $\Delta_p\ll \Delta_{0p}$\cite{burin04},
 \bea\no
\label{eq.1}
\frac{da_1}{dt}&=&\delta_0e^{i\delta t}a_2(t),\\
\frac{da_2}{dt}&=&-\delta_oe^{-i\delta t}a_1(t),
\eea
where $\delta=\omega-\frac{E}{\hbar}$,
$\delta_0=\frac{\xi_0 p_{12}}{2 \hbar}$ and
$p_{12}=p_0\frac{\Delta_{0p}}{E}$ in which $p_0$ is the dipole moment
of the ETLS. $E=E_2-E_1$ where $E_1$ and $E_2$ are the energy of
ground and the exited states of the ETLS, respectively.
We can easily find that at $t=\tau_1$ the coefficients are given by,
\bea\label{eq.2}\no
a_1(\tau_1)&=&\frac{a_1(0)\gamma_2 -i\delta_0
a_2(0)}{\gamma_1+\gamma_2}e^{i\gamma_1\tau_1}
\\\no&&+\frac{a_1(0)\gamma_1
+i\delta_0 a_2(0)}{\gamma_1+\gamma_2}e^{-i\gamma_2\tau_1},\\\no
a_2(\tau_1)&=&\frac{i}{\delta_0}[\frac{a_1(0)\gamma_2 -i\delta_0
a_2(0)}{\gamma_1+\gamma_2}\gamma_1e^{i\gamma_2\tau_1}
\\&&-\frac{a_1(0)\gamma_1
+i\delta_0
a_2(0)}{\gamma_1+\gamma_2}\gamma_2e^{-i\gamma_1\tau_1}],
\eea
where
$\gamma_1=\frac{\delta+\sqrt{\delta^2+4\delta_0^2}}{2}$ and
$\gamma_2=\frac{-\delta+\sqrt{\delta^2+4\delta_0^2}}{2}$.
The initial condition is imposed by the Boltzmann
weight, i.e.
$|a_1(0)|^2=\frac{1}{Z}e^{ -\beta E_1}$ and
$|a_2(0)|^2=\frac{1}{Z}e^{ -\beta E_2}$, where  $Z=e^{ -\beta
E_1}+e^{ -\beta E_2}$  is the partition function,
$\beta=\frac{1}{K_B T}$ and $K_B$ is the Boltzmann constant.

The system  relaxes for a time duration of $t_0$ before the
second pulse arrives. The second pulse of the electric field tries to
recover the initial configuration during $\tau_2=2\tau_1$ seconds. Finally
the measurement on the amplitude of the electric dipole is done
at $t'=t_0$ after switching off the second pulse. This can be
expressed by the following equation,
 \bea
 \no A(2t_0)&=&<\psi(2t_0)\mid p\mid\psi(2t_0)>
 \\&\simeq&
 2Re[a_1^{*}(\tau_2)a_2(\tau_2)p_{12}e^{-\frac{iEt_0}{\hbar}}]
 \label{echo-amplitude}
 \eea
where $p=p_0\sigma_{z}$ and the details of the other coefficients
can be found in the appendix.

As mentioned before we will consider the effect of the magnetic
field via a Mexican hat model. In this
model for describing the non-monotonic magnetic field dependence
of the dielectric susceptibility of multi-component glasses,
Kettemann et al.\cite{kettemann99} investigated the properties of
tunneling systems exhibiting the peculiarity that the tunneling
particle can move along different paths to go from one potential
minimum to the other, thus forming a closed tunneling loop. As a
result in the three dimensional Mexican hat model, using the
Aharanov-Bohm effect and the assumption that the tunneling through
both paths occurs with equal probability, the tunnel splitting
becomes a periodic function of the magnetic flux threading through
the loop,
\begin{equation}
\Delta_{0p}\rightarrow \Delta_{0p}(\phi)=\Delta_{0p}\cos(\pi\frac{\phi}{\phi_{0}}),
\label{Delta-phi}
\end{equation}
where $\phi$ is the total magnetic flux passing through a closed
tunneling path, $\phi_0$ is the renormalized quantum flux defined by
$\phi_0=\frac{h}{q}$ in the previous section.
%%%%%%%%%%%%%%%%%%%%%%%%%%%%%%%%%%%%%%%%%%%%%%%%%%%%%%%%%%%%%%%%%%%%
%%%%%%%%%%%%%%%%%%%%%%%%%%%%%%%%%%%%%%%%%%%%  figure 3
\begin{figure}
\includegraphics[width=\columnwidth]{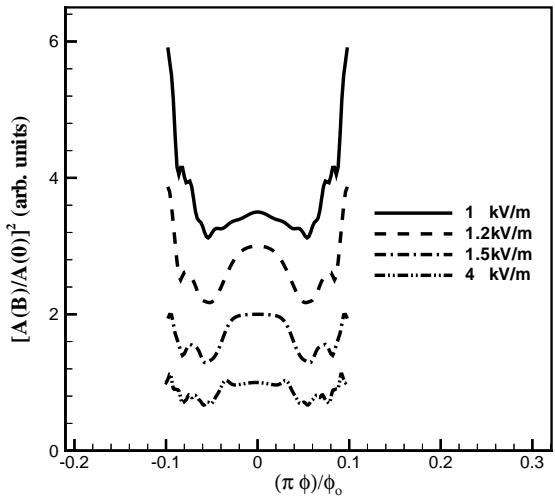}
\caption{ The echo amplitude versus the magnetic flux ratio for
different electric field amplitudes, by averaging over the effective
distribution function for $0<\frac{\Delta}{\hbar}<10^8 Hz $ and
$10^9<\frac{\Delta_o}{2\hbar}<10^9+10^6 Hz$, ($\omega=2\times 10^9
Hz$, $t_0=2\mu s$ and $T=15mK$). The vertical axis has been
shifted.} \label{fig3}
\end{figure}
%%%%%%%%%%%%%%%%%%%%%%%%%%%%%%%%%%%%%%%%%%%%%%%%%%%%%%%%%%%%%%%%%%%%%
%%%%%%%%%%%%%%%%%%%%%%%%%%%%%%%%%%%%%%%%%%%%%%%%%%%%%%%%%%%%%%%%%%%%
%%%%%%%%%%%%%%%%%%%%%%%%%%%%%%%%%%%%%%%%%%%%  figure 4
\begin{figure}
\includegraphics[width=\columnwidth]{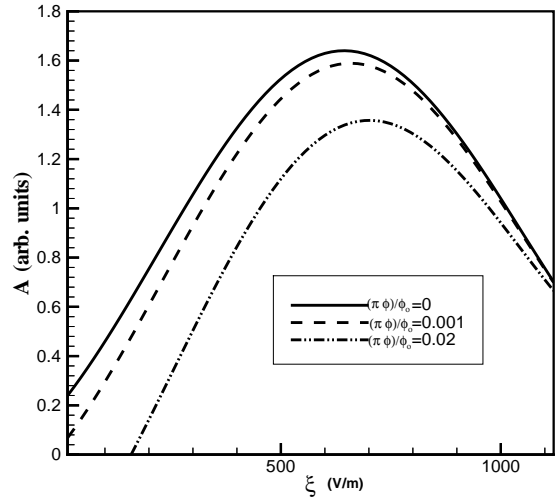}\caption{
The echo amplitude versus electric field amplitude, for different
values of the
magnetic field flux,  by averaging over the effective distribution
function for $0<\frac{\Delta}{2\hbar}<10^7 Hz $ and
$10^9<\frac{\Delta_o}{2\hbar}<10^9+10^6 Hz $,  ($\omega=2\times
10^9 Hz$,  $t_0=1\mu s$ and $T=15mK$).} \label{fig4}
\end{figure}
%%%%%%%%%%%%%%%%%%%%%%%%%%%%%%%%%%%%%%%%%%%%%%%%%%%%%%%%%%%%%%%%%%%%%
%%%%%%%%%%%%%%%%%%%%%%%%%%%%%%%%%%%%%%%%%%%%%%%%%%%%%%%%%%%%%%%%%%%%

\section{Results on echo amplitude}
\label{results}

In this section we are going to present our results on the
amplitude of the echo response (Eq.(\ref{echo-amplitude})) as a
function of the applied magnetic and electric field in addition to
the pulse separation time ($t_0$) and temperature.

In the absence  of a magnetic field  the echo amplitude measured
for small pulse separations in an ideal double-pulse experiment,
ignoring relaxation, has a maximum when the amplitude of the driving
electric field is given by\cite{phill_87} \bea
\frac{p_{12}\xi_0\tau_1}{2\hbar}=\frac{\pi}{4} \label{eq.5} \eea
and at $T=10 mK$ the tunneling matrix element is
$\frac{\Delta_0}{\hbar}\sim 10^9 Hz$. If we consider the typical value
for $\xi_o=1\frac{kV}{m}$ and  $\tau_1= 100ns$ then
$\frac{p_{12}\xi_o}{2\hbar}\sim 10^6-10^7Hz$. To compare our
results with recent
experiments\cite{ludwig02-a,ludwig02-b,ludwig03,nagel04} the
following frequency $\frac{\omega}{2\pi}\sim 10^{8}-10^9 Hz$ and
pulse separation $t=t_o\sim 10^{-6}s$ are assumed.

We should mention that the result of Eq.(\ref{echo-amplitude}) is valid
for low magnetic field amplitudes. At high magnetic fields the effective
tunneling amplitude $\Delta_{0p}\cos(\pi\frac{\phi}{\phi_{0}})$ becomes small
and the resonance condition $(|\delta|\ll\omega)$ is not fulfilled.
Therefore, the resonance condition
($|\omega-\frac{\sqrt{\Delta_p^2+\Delta_{0p}(\phi)^2}}{\hbar}|\ll \omega$)
is satisfied whenever
$1-\cos(\pi \frac{\phi}{\phi_{0}})\ll 1$, having assumed that
$\frac{\Delta_{0p}}{\hbar}\sim \omega$.
In other words the resonance condition
is guaranteed for $(|\pi \frac{\phi}{\phi_{0}}|\ll 1)$.

\subsection{Magnetic field dependence}

We have plotted in Fig.(\ref{fig2}), the square of the normalized echo
amplitude ($|A(B)/A(B=0)|^2$) versus the magnetic field. The magnetic
field ($B$) in the horizontal axis is presented as flux
($\phi=SB$, $S$ is the effective area of ETLS) divided by the quantum
flux ($\phi_0$) to simplify data analysis. To arrive at this plot
we have used Eq.(\ref{echo-amplitude}) with $t_0=2\mu s$. We have
also averaged over the tunneling parameters ($\Delta_{p},
\Delta_{0p}$). Note that, in this process we have assumed
$\Delta_{p}<\Delta_{0p}$ to have reasonable flip-flop
configurations \cite{burin04,burin98}. Moreover the resonance
condition of the ETLS limits the range of $\Delta_{0p}$ to be very
close to the frequency of the electric field
$10^9<\frac{\Delta_{0p}}{2\hbar}<10^9+10^6 Hz$. The distribution
of the ETLS parameters is given by Eq.(\ref{f2}). The echo
amplitude has a maximum at $B=0$ as presented in Fig.(\ref{fig2}).
It then decreases to a minimum before rising up. This is in
agreement with the experimental observation presented in
Ref.[\onlinecite{ludwig02-a,ludwig02-b,ludwig03,nagel04}].

\subsection{Dependence on the electric field}

We have examined the dependence of our results on the
amplitude of the electric field. The echo amplitude versus the magnetic
field for different strengths of the electric field
is plotted in Fig.(\ref{fig3}).
The data for each value of the electric field have been shifted by a constant
for clarity. We will see that the height of the central peak
is enhanced by increasing the electric field. This feature also coincides
with the experimental data on the defected crystal $KBr:CN$
(Fig.1 in Ref.[\onlinecite{ludwig02-b}]).

We have then plotted the echo amplitude versus the amplitude of
the electric field in Fig.(\ref{fig4}) for different magnetic fields.
The echo amplitude rises and has a maximum for an electric field
around $700 V/m$, it then decreases for larger fields. The general feature
is in agreement with the experiments on multicomponent glasses
(Fig.2 in Ref.[\onlinecite{ludwig02-a}]).

%%%%%%%%%%%%%%%%%%%%%%%%%%%%%%%%%%%%%%%%%%%%%%%  figure 5
\begin{figure}
\includegraphics[width=\columnwidth]{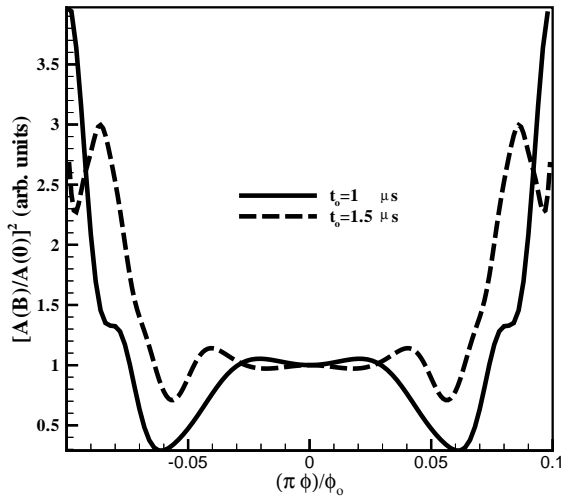}
\caption{ The echo amplitude versus the magnetic flux ratio for
different delay times  $t_o=1\mu s $ and $t_o=1.5\mu s $, by
averaging over the effective distribution function for
$0<\frac{\Delta}{\hbar}<10^8 Hz $ and
$10^9<\frac{\Delta_o}{2\hbar}<10^9+10^6 Hz $, ($\omega=2\times
10^9 Hz$, $\frac{p_o\xi_o}{2\hbar}=7\times 10^6 Hz $ and
$T=15mK$).} \label{fig5}
\end{figure}
%%%%%%%%%%%%%%%%%%%%%%%%%%%%%%%%%%%%%%%%%%%%%%%%%%%%%%%%%%%%%%%%%%%%%%%
%%%%%%%%%%%%%%%%%%%%%%%%%%%%%%%%%%%%%%%%%%%%%%%  figure 6
\begin{figure}
\includegraphics[width=\columnwidth]{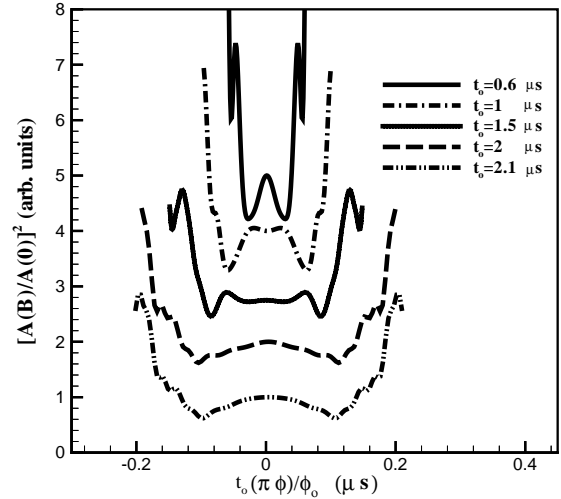}
\caption{ The echo amplitude versus the magnetic flux ratio for
different delay times, by averaging over the effective distribution
function for $0<\frac{\Delta}{\hbar}<10^8 Hz $ and
$10^9<\frac{\Delta_o}{2\hbar}<10^9+10^6 Hz $, ($\omega=2\times
10^9 Hz$, $\frac{p_o\xi_o}{2\hbar}=7\times 10^6 Hz $ and
$T=15mK$). The vertical axis has been shifted.} \label{fig6}
\end{figure}
%%%%%%%%%%%%%%%%%%%%%%%%%%%%%%%%%%%%%%%%%%%%%%%%%%%%%%%%%%%%%%%%%%%%%%%

\subsection{Effect of pulse separation}

We have presented in Fig.(\ref{fig5}) the same quantity as Fig.(\ref{fig2})
but for different pulse separations ($t_0=1, 1.5 \mu s$). Although the global
behavior is similar to Fig.(\ref{fig2}) there is a slight difference at
small magnetic fields. We have observed that for short pulse separations,
$t_0=1\mu s$,
the echo amplitude gets a minimum at $B=0$. The effect is similar to the
experiments on crystals with point defects  (Fig.2 of Ref.[\onlinecite{ludwig02-b}]).
For short pulses the magnetic field does not have a dephasing effect.
However the central peak broadening does not accompany the decreasing of
pulse separation as in Fig.3 of Ref.[\onlinecite{ludwig02-b}]. In contrast,
the broadening of the central peak decreases with $t_0$. This can be seen
in Fig.(\ref{fig6}) where we have scaled the horizontal axis by a factor
of $t_0$. The width of the central peak diminishes by applying
shorter pulse separations.

\subsection{Temperature dependence}

\begin{figure}
\includegraphics[width=\columnwidth]{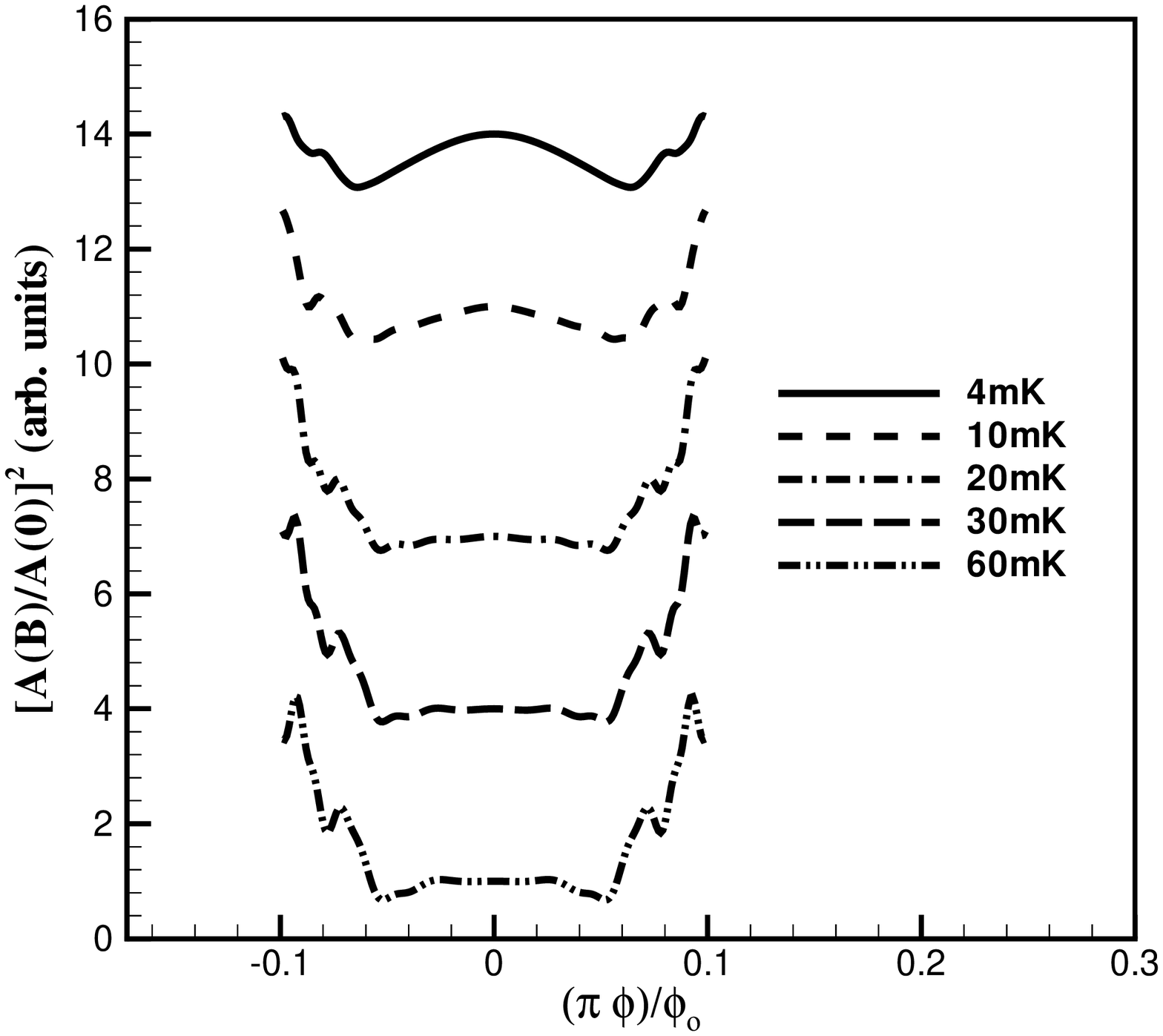}\caption{
The echo amplitude versus the magnetic flux ratio for Different
temperatures, by averaging over effective distribution function
for $0<\frac{\Delta}{\hbar}<10^8 Hz $ and
$10^9<\frac{\Delta_o}{2\hbar}<10^9+10^6 Hz $, ($\omega=2\times
10^9 Hz$, $t_o=2\mu s $ and $\frac{p_o\xi_o}{2\hbar}=7\times 10^6
Hz $). The vertical axis has  been shifted.} \label{fig7}
\end{figure}

The temperature dependence of our results enters via the Boltzmann weight
as the initial probability of the states in ETLS. This dependence is
plotted in Fig.(\ref{fig7}). We have plotted the amplitude of the
echo response
versus magnetic field for different temperatures. To have a clear plot for
different curves we have added a constant value to each curve. Thus they
are well represented instead of falling onto each other. We have seen that the
intensity of the central peak is reduced by increasing the
temperature and becomes
almost flat at high temperatures. Apart from that the general behavior is
unchanged. This can also be compared to the experimental data on the multicomponent
glasses (Fig.8 of Ref.[\onlinecite{ludwig03}). Although the profile of the
echo amplitude versus magnetic field does not fit completely, the reduction of central peak
is in agreement.
%%%%%%%%%%%%%%%%%%%%%%%%%%%%%%%%%%%%%%%%%%%%%%%%%%%%%%%%%%%%%%%%%%%%%%%
%%%%%%%%%%%%%%%%%%%%%%%%%%%%%%%%%%%%%%%%%%%%%%%  figure 7
%%%%%%%%%%%%%%%%%%%%%%%%%%%%%%%%%%%%%%%%%%%%%%%%%%%%%%%%%%%%%%%%%%%%%%%

\section{Summary and discussion}

We have presented a model to explain the anomalous behavior of
the echo response in glasses at low temperatures. This model is essentially
based on an effective TLS which is the result of interaction between the
original TLSs in glasses or defected crystals. Most of the experimental features at
low magnetic fields can be explained by our approach. In this respect, the dephasing
effect which reduces the echo amplitude at low fields (Fig.(\ref{fig2})),
the dependence on the strength of the electric field (Figs.(\ref{fig3}, \ref{fig4}))
and also the effect of various temperatures (Fig.(\ref{fig7}))
have been addressed.
However, in the response
of different pulse separations ($t_0$) although
it shows the broadening of the central peak
by reduction of $t_0$ the product of $t_0 \phi$ does not seem to be universal which
deviates from the experimental data.

As we have already mentioned our model is valid for the low magnetic field regime.
At high magnetic fields the resonance condition is lost, thus the flip-flop
basis of ETLS does not contribute effectively.
Because, if a TLS is initially in the ground state,
the probability to find the system in the second
state is given
by\cite{burin98}
\bea
W_{12}(t)=\frac{\Delta_0(\phi)^2}{\Delta^2+\Delta_0(\phi)^2}
\sin^2(\frac{\sqrt{\Delta^2+\Delta_0(\phi)^2}\cdot t}{2\hbar}).
\eea
For high magnetic fields
this probability vanishes since
$\Delta_0(\phi)$ become very small. So the particles will
be frozen in the lowest well and do not respond to the external
electric field. Thus we expect the echo amplitude for high
magnetic fields to be constant.

The other aspect of echo response which can be explained by this model is
the response to the frequency of the electric field. It has been shown
that the echo amplitude of $BK7$ is almost independent of electric field
frequency (see Fig.(7) in Ref.[\onlinecite{ludwig03}]). A raw description
is given by taking into account the interaction between ETLS defined in
Sec.\ref{etls}. Thus the resonance condition can take place for different
frequencies and the final response will be the same. Note that the distribution
of the effective ETLS is given by Eq.(\ref{f2}) which is given
in the appendix.

However, there is a debate on the interpretation of the results which
comes from a Mexican hat model. Let us reduce the discussion to two
points. The first point is related to the route where the magnetic field
enters the problem. In our model (Mexican hat type) the magnetic field effect
appears in the tunneling amplitude (Eq.(\ref{Delta-phi})) via the Aharanov-Bohm
phenomenon. This by itself can not be ruled out. Recently, Cochec, et al. \cite{cochec02}
reported experiments on the dielectric constant of a structural glass and
explained their results by considering the variation of the tunneling amplitude
versus magnetic field. This is based on a microscopic theory of transport in
disordered networks which can be interpreted as the tunneling between two wells.
The effect of the magnetic field due to quantum phases
influences the tunneling amplitude \cite{medina92}.

The second point is related to the interpretation of a large
effective charge $q$, (or small quantum flux $\phi_0$) which
scales the magnetic field to the right value of the experimental
data. This is the most controversial point and we are going to add
some comments in this respect. Firstly, the large value of $q$ has
been taken as an indication that there is a  coherent motion of
tunneling particles taking place as the result of their mutual
interaction\cite{kettemann99,langari02}. As far as the low
temperature physics of glasses is concerned the interaction
between TLSs is an important issue which has been recently
addressed as a dephasing mechanism in the relaxation of cold
glasses \cite{burin04}. The relaxation can be enhanced by
interaction through flip-flop configurations leading to
delocalized state. Thus the interaction should also have a
considerable effect in the dephasing mechanism in the presence of
a magnetic field. And this is taken into account in the Mexican
hat type model effectively. Secondly, the nature of a tunneling
system (TLS) is not clearly known. The general belief is that it
represents the low energy excitations of glasses at low
temperatures. It has been recently argued that such types of
excitation come from two energetically close configurations of a
cluster of atoms or molecules \cite{lubchenko2001, lubchenko04}.
The cluster contains roughly $N=200$ atoms or molecules
contributing in a collective tunneling. Thus the flux passing
through such a cluster is much larger than the case in a single
atom. This by itself can improve the necessary flux ($\phi$) in
the Mexican hat model by two orders of magnitude. Taking into
account the interaction between the mentioned tunneling systems
gives the right order of the magnetic field.

It should be cited that due to the new experiments on glasses
with different isotopes the effect of the nuclear quadrupole moment (NQM) is of
significant importance \cite{nagel04} which is lacking in our model.
The theory which is based on NQM
works with a single molecule as a tunneling center
and takes its NQM into account \cite{wfe02}.
However, the next step is to introduce
an effective nuclear quadrupole moment for a tunneling system which is
a cluster of molecules.
The theory explains very well
the low and hight magnetic field behavior of the echo response.
But the intermediate
regime is missing.
Besides, in this approach the static electric susceptibility
is six orders of magnitude smaller than the experimental one.
We should mention that our model gives the  temperature dependence
of the echo amplitude in the presence of a  magnetic field
which is absent in the
NQM model proposed in Ref.[\onlinecite{wfe02}].

%%%%%%%%%%%%%%%%%%%%%%%%%%%%%%%%%%%%%%%%%%%%%%%%%%%%%%%%%%%%%%%%%%%%%%%%%%%%

\acknowledgments
We would like to express our deep gratitude to
P. Fulde for valuable comments and useful discussions.
We would also like to thank D. Bodea and I. Ya. Polishchuk
for fruitful discussions.

%%%%%%%%%%%%%%%%%%%%%%%%%%%%%%%%%%%%%%%%%%%%%%%%%%%%%%%%%%%%%%%%%%%%%%%%%%%%
%%%%%%%%%%%%%%%%%%%%%%%%         Appendix       %%%%%%%%%%%%%%%%%%%%%%%%%%%%
\begin{appendix}
\section{The echo amplitude after the second pulse}

As mentioned in the text
the Boltzmann weight is used to define the initial condition,
$|a_1(0)|^2=\frac{1}{Z}e^{ -\beta E_1}$ and
$|a_2(0)|^2=\frac{1}{Z}e^{ -\beta E_2}$, where  $Z=e^{ -\beta
E_1}+e^{ -\beta E_2}$  is the partition function,
$\beta=\frac{1}{K_B T}$ and $K_B$ is the Boltzmann constant. So the
wave function after the relaxation time $t_0$ is
\bea\no\mid
\psi(t_0)\rangle&=&
a_1(\tau_1)e^{\frac{-iE_1(t_0+\tau_1)}{\hbar}}\mid\varphi_1\rangle+\\\no
&&
a_2(\tau_1)e^{\frac{-iE_2(t_0+\tau_1)}{\hbar}}\mid\varphi_2\rangle.
\eea

During the second  pulse the probability  amplitudes  must satisfy
Eq.~(\ref{eq.1}), so in a time $t^{\prime}$ after the  second
pulse we have

\bea\no\mid
\psi(t_0+t^{\prime})\rangle=&&
a_1(\tau_2)e^{\frac{-iE_1t^{\prime}}{\hbar}}\mid\varphi_1\rangle+
\\\no&&a_2(\tau_2)e^{\frac{-iE_2t^{\prime}}{\hbar}}\mid\varphi_2\rangle,
\eea
where
\bea\label{eq.3}\no
a_1(\tau_2)&=&\frac{a_1^{\prime}(\tau_1)\gamma_2 -i\delta_0
a_2^{\prime}(\tau_1)}{\gamma_1+\gamma_2}e^{i\gamma_1\tau_2}
\\\no&&+\frac{a_1^{\prime}(\tau_1)\gamma_1
+i\delta_0
a_2^{\prime}(\tau_1)}{\gamma_1+\gamma_2}e^{-i\gamma_2\tau_2},\\\no
a_2(\tau_2)&=&\frac{i}{\delta_0}[\frac{a_1^{\prime}(\tau_1)\gamma_2
-i\delta_0
a_2^{\prime}(\tau_1)}{\gamma_1+\gamma_2}\gamma_1e^{i\gamma_2\tau_2}
\\\no&&-\frac{a_1^{\prime}(\tau_1)\gamma_1
+i\delta_0
a_2^{\prime}(\tau_1)}{\gamma_1+\gamma_2}\gamma_2e^{-i\gamma_1\tau_2}],
\eea
\bea \no
a_1^{\prime}(\tau_1)&=&a_1(\tau_1)e^{\frac{-iE_1(t_0+\tau_1)}{\hbar}},\\
a_2^{\prime}(\tau_1)&=&a_2(\tau_1)e^{\frac{-iE_2(t_0+\tau_1)}{\hbar}}.
\eea
By these assumptions,  the amplitude of the two pulse echo from an effective
TLS at time $2t_0$ measured from the first pulse is given by
 \bea
 \no A(2t_0)&=&\langle\psi(2t_0)\mid p\mid\psi(2t_0)\rangle\\&\simeq&
 2Re[a_1^{*}(\tau_2)a_2(\tau_2)p_{12}e^{-\frac{iEt_0}{\hbar}}],
 \eea
where $p=p_0\sigma_{z}$ and $\tau_2=2\tau_1$.

\section{The distribution of tunneling parameters
in coupled interacting ETLSs}

If we consider that the interaction between two ETLSs is the same as the
interaction between the original TLSs
then we can  use the same procedure used previously for a pair of
TLSs. From Eq.(\ref{eq.6}) we have
\bea
\no f^{(3)}(\Delta_p,\Delta_{0p})\propto T^2
\int\frac{d\Delta_{01}}{\Delta_{01}^2}\int\frac{dE_1E_1}{\sqrt{E_1^2-\Delta_{01}^{
2}}}n(E_1)\times\\
\no\int\frac{d\Delta_{02}}{\Delta_{02}^2}
\int\frac{dE_2E_2}{\sqrt{E_2^2-\Delta_{02}^{2}}}[1-n(E_2)]\times\\
\int d^3R\delta(\Delta_p-\mid E_1-E_2\mid)
\delta(\Delta_{0p}-\frac{U_0}{R^3}\frac{\Delta_{01}\Delta_{02}}{E_1E_2}),
\no
\eea
where $n(E_1)=[1+\exp(\beta E_1)]^{-1}$ is the probability of
finding an ETLS in the exited state. By  a little manipulation we can
show that
\bea
f^{(3)}(\Delta_p,\Delta_{0p})\propto
\frac{T}{\Delta_{0p}^2}[\ln(\frac{2T}{\Delta_{0min}})]^2.
\eea
Since $\Delta_{0min}\sim T$, we   find that  the distribution
function for the coupled ETLS parameters is similar to the second
distribution function (Eq.(\ref{f2})).

\end{appendix}

%%%%%%%%%%%%%%%%%%%%%%%%%%%%%   References

\end{document}